\documentclass{article}
\begin{document}

\pagenumbering{arabic}
\title{Physics of dark energy particles}
\author{
  C. G. B\"ohmer\footnote{E-mail: c.boehmer@ucl.ac.uk}\\
  Department of Mathematics, University College London,\\
  Gower Street, London, WC1E 6BT, UK
\and 
  T. Harko\footnote{E-mail: harko@hkucc.hku.hk}\\
  Department of Physics and Center for Theoretical\\
  and Computational Physics, The University of Hong Kong,\\
  Pok Fu Lam Road, Hong Kong
}
\date{}
\maketitle

\begin{abstract}
We consider the astrophysical and cosmological implications of the
existence of a minimum density and mass due to the presence of the
cosmological constant. If there is a minimum length in nature,
then there is an absolute minimum mass corresponding to a
hypothetical particle with radius of the order of the Planck
length. On the other hand, quantum mechanical considerations
suggest a different minimum mass. These particles associated with
the dark energy can be interpreted as the ``quanta'' of the
cosmological constant. We study the possibility that these
particles can form stable stellar-type configurations through
gravitational condensation, and their Jeans and Chandrasekhar
masses are estimated. From the requirement of the energetic
stability of the minimum density configuration on a macroscopic
scale one obtains a mass of the order of $10^{55}{\rm g}$, of the
same order of magnitude as the mass of the universe. This mass can
also be interpreted as the Jeans mass of the dark energy fluid.
Furthermore we present a representation of the cosmological
constant and of the total mass of the universe in terms of
`classical' fundamental constants.
\end{abstract}

\mbox{} \\ 
\mbox{} 
\mbox{} \\
\mbox{} 

\newpage
\section{Introduction} \label{intro}
Several recent astrophysical observations of distant type Ia
supernovae \cite{Ri98,Pe99,Ber00,Ha00} have provided the
astonishing result that around $95-96\%$ of the content of the
universe is in the form of dark matter $+$ energy, with only about
$4-5\%$ being represented by baryonic matter. More intriguing,
around $70\%$ of the total energy-density may be in the form of
what is called the dark energy, with the associated density parameter $%
\Omega _{DE}$ of the order of $\Omega _{DE}\sim 0.70$. The dark
energy is responsible for the recent acceleration of the Universe.
The best candidate for the dark energy is the cosmological
constant $\Lambda $, which is usually interpreted physically as a
vacuum energy, with energy density $\rho _{\Lambda }$ and pressure
$p_{\Lambda }$ satisfying the unusual equation of state $\rho
_{\Lambda }=-p_{\Lambda }/c^{2}=\Lambda /8\pi G/c^2$. Its size is of
the order $\Lambda \approx 3\times 10^{-56}$
cm$^{-2}$~\cite{PeRa03,Pa03}.

The existence of the cosmological constant modifies the allowed
ranges for various physical parameters, like, for example, the
maximum mass of compact stellar
objects~\cite{MaDoHa00,Bo04,Bo05,BaBoNo05a}, thus leading to a
modifications of the ``classical'' Buchdahl limit~\cite{Bu}. In
conjunction with other parameters, like the Schwarzschild radius,
the cosmological constant $\Lambda $ leads to a set of scales
relevant not only for cosmological, but also for astrophysical
applications. Hence, for example, there exists a lower and an
upper cut-off on the possible velocities of test particles
travelling over distances of the order of Mpc~\cite{BaBoNo05}.

Since about $70\%$ of the Universe consists of dark energy, which
almost
entirely determines its structure and dynamics, it is natural to consider $%
\Lambda $ as a fundamental constant and to explore the
possibilities which follows from enlarging the set of fundamental
constants of nature, which can
be considered as being the speed of light $c$, the gravitational constant $G$%
, Planck's constant $\hbar $ and the cosmological constant
$\Lambda $, respectively~\cite{We04}. On the other hand, we cannot
exclude \textit{a priori} the possibility that the cosmological
constant, which may also be interpreted as a manifestation of the
vacuum energy, can also play an important role not only at
galactic or cosmological scales, but also at the level of
elementary particles. Therefore the presence of the cosmological
constant may require a drastic modification of the basic laws of
physics.

In the presence of a cosmological constant, ordinary Poincar\'{e}
special relativity is no longer valid, and must be replaced by a
de Sitter special relativity, in which Minkowski space is replaced
by a de Sitter spacetime \cite{Pe1}. Consequently, the ordinary
notions of energy and momentum change, and will satisfy a
different kinematic relation.  Since the only difference between
the Poincar\'{e} and the de Sitter groups is the replacement of
translations by certain linear combinations of translations and
proper conformal transformations, the net result of this change is
ultimately the breakdown of ordinary translational invariance
\cite{Pe2,Pe3,Pe4}. From the experimental point of view,
therefore, a de Sitter special relativity might be probed by
looking for possible violations of translational invariance. If we
assume the existence of a connection between the energy scale of
an experiment and the local value of the cosmological constant,
there would be changes in the kinematics of massive particles
which could hopefully be detected in high-energy experiments.
Furthermore, due to the presence of a horizon, the usual causal
structure of spacetime would be significantly modified at the
Planck scale.

By using dimensional analysis, Wesson~\cite{We04} has found two
different masses, which can be constructed from the set of constants $%
(c,G,\hbar ,\Lambda )$. The mass $m_{P}$ relevant at the quantum
scale is
\begin{equation}
m_{P}=\left( \frac{\hbar }{c}\right) \sqrt{\frac{\Lambda
}{3}}\approx 3.5\times 10^{-66}\,\mathrm{g},  \label{mp}
\end{equation}
while the mass $m_{PE}$ relevant to the cosmological scale is
\begin{equation}
m_{PE}=\left( \frac{c^{2}}{G}\right) \sqrt{\frac{3}{\Lambda
}}\approx 1\times 10^{56}\,\mathrm{g}.  \label{mpe}
\end{equation}

The interpretation of the mass $m_{PE}$ is straightforward: it is
the mass of the observable part of the universe, equivalent to
$10^{80}$ baryons of mass $10^{-24}$ g each. The interpretation of
the mass $m_{P}$ is more difficult. By using the dimensional
reduction from higher dimensional relativity and by assuming that
the Compton wavelength of a particle cannot take any value,
Wesson~\cite{We04} proposed that the mass is quantised according
to the rule $m=(n\hbar /c)\sqrt{\Lambda /3}$. Hence $m_{P}$ is the
minimum mass corresponding to the ground state $n=1$.

With the use of the generalized Buchdahl identity~\cite{MaDoHa00},
it can be rigorously proven that the existence of a non-negative
$\Lambda $ imposes a lower bound on the mass $M$ and density $\rho
$ for general relativistic objects with radius $R$, which is given
by~\cite{BoHa05}
\begin{equation}
2GM\geq \frac{\Lambda c^{2}}{6}R^{3},\qquad \rho =\frac{3M}{4\pi
R^{3}}\geq \frac{\Lambda c^{2}}{16\pi G}=:\rho _{\min }.
\label{minm}
\end{equation}

Therefore, the existence of the cosmological constant implies the
existence of an absolute minimum density in the universe.
No object present in relativity can have a density that is smaller
than $\rho_{\min}$. For $\Lambda >0$, this result also implies a
minimum density for stable fluctuations in energy density. These
results have been generalized to compact anisotropic general
relativistic objects in \cite{Bo1}, where it was shown that in the
presence of the cosmological constant, a minimum mass
configuration with given anisotropy does exist. For charged
general relativistic objects there is also a lower bound for the
mass-radius ratio \cite{Bo2}. By considering the total energy
(including the gravitational one) and the stability of the objects
with minimum mass-radius ratio,  a representation of the mass and
radius of the charged objects with minimum mass-radius ratio in
terms of the charge and vacuum energy only has been obtained.

It is the purpose of the present paper to further explore the
possible physical implications of the existence of a minimum mass
in the universe, given by Eq.~(\ref{minm}), and which is a direct
consequence of the existence of a non-zero cosmological constant.
In particular, we show that if a minimum length does exist in
nature, then the condition~(\ref{minm}) does imply the existence
of an absolute minimum mass. By combining the rigorous result for
the minimum mass with the dimensional analysis of Wesson~
\cite{We04}, we can obtain an intriguing representation of the
vacuum energy as a function of the fundamental constants $c$, $G$,
$\hbar $ as well as the mass $m_{e}$ and the radius $r_{e}$ of the
electron. On the other hand, by considering the possibility of the
gravitational condensation of the dark energy fluid we obtain the
interpretation of the mass $m_{PE}$ as the Jeans mass of the
gravitational dark energy condensate. By minimizing the total
(matter plus gravitational) energy of a stable configuration
consisting of particles with the minimum mass we provide a
rigorous derivation of the cosmological mass $m_{PE}$, given by
Eq.~(\ref{mpe}).

The present paper is organized as follows. The physical
implications of the existence of a minimum mass are presented in
Section 2. The gravitational condensation of the dark energy
particles is considered in Section 3. The total energy (including
the gravitational one) for stable configurations of particles with
minimum mass is obtained in Section 4. We briefly conclude and
discuss our results in the last section.

\section{Minimum mass and radius of dark energy particles}

At a microscopic level two basic quantities, the Planck mass
$m_{Pl}$ and the Planck length $l_{Pl}$ are supposed to play a
fundamental physical role. The Planck mass is derived by equating
the gravitational radius $2Gm/c^{2}$ of a Schwarzschild mass with
its Compton wavelength $\hbar /mc$. The
corresponding mass $m_{Pl}=\left( c\hbar /2G\right) ^{1/2}$ is of the order $%
m_{Pl}\approx 1.5\times 10^{-5}$ g. The Planck length is given by $%
l_{Pl}=\left( \hbar G/c^{3}\right) ^{1/2}\approx 1.6\times
10^{-33}$ cm and at about this scale quantum gravity will become
important for understanding physics. The Planck mass and length
are the only parameters with dimension mass and length,
respectively, which can be obtained from the fundamental constants
$c$, $G$ and $\hbar $.

The problem of the physical nature of the cosmological
constant/dark energy is one of the most important issues
confronting modern physics. A popular interpretation of the
cosmological constant is in terms of the vacuum energy $\langle\rho_{vac}\rangle$,
which is of the order $\langle\rho _{vac}\rangle\approx 2\times
10^{71}$ GeV$^4$. However, astronomical observations indicate that
the cosmological constant is many orders of magnitude (around
$10^{120}$) smaller than the estimate for vacuum energy. Many
different approaches to the solution of this problem have been
proposed, like the interpretation of the cosmological constant as
an integration constant, anthropic considerations, quantum
cosmology etc. \cite{We89}. Presently, astronomical observations
suggests that dark energy could be dynamical and evolving, with
the dark-energy density approaching its natural value, zero. The
smallness of the dark energy is a result of the expansion of the
universe and its old age \cite{PeRa03}.

Due to the fact that in a curved space-time the vacuum is not
unique, the phenomenon of particle production occurs in an
expanding universe as a typical quantum effect \cite{BiDa82}. As
the universe evolves, and the curvature changes, the vacuum state
also changes, and the initial zero particle vacuum state becomes
later a multiparticle state. If the universe is decelerating and
is asymptotically Minkowskian at infinity, in the large time limit
the particle production does not occur any more. However,
observations indicate that we are living in an accelerating
universe, and the mechanism of particle production can be very
important. The rate of the particle production for a flat universe
filled with a fluid with equation of state $p=\alpha \rho$, where
$\alpha $ is arbitrary, has been obtained recently in
\cite{BaFaHo07}. The calculations were performed for the case of a
massless scalar field, for which the corresponding Klein-Gordon
equation was solved. The rate of particle production is determined
exactly for any value of $\alpha $, including $\alpha =-1$. When
the strong energy condition is satisfied, the rate of particle
production decreases as time goes on, in agreement to the fact
that the four-dimensional curvature decreases with the expansion;
the opposite occurs when the strong energy condition is violated.
Hence the "cosmological constant" (with $\rho c^2=-p$) can be an
effective source of particles, during a purely de Sitter
evolutionary phase \cite{BaFaHo07}.

An alternative approach to particle production from a dark energy
vacuum fluid was suggested in \cite{We05}, by assuming that the
vacuum, with the energy density proportional to $\Lambda $, gives
up energy which corresponds to a particle with rest mass $m$, so
that $d\Lambda =-6\left(mc/h\right)^2$. This situation is similar
to the Dirac hole theory, in which a positron is regarded as a
hole created in an underlying sea of energy. Particle production
from dark energy can also be interpreted in geometrical terms. The
vacuum energy/cosmological constant is a "sea of energy" that
curves the space-time having a curvature $L=\sqrt{3/\Lambda }$.
Locally, a perturbation in the vacuum corresponds to a change in
curvature, and a change in the curvature leads to a change in the
quantum mechanical vacuum state, resulting in a production of a
massive particle.

>From Eq.~(\ref{minm}) one can estimate the numerical value of the
minimal density for a positive $\Lambda $ as $\rho _{\min
}=\Lambda c^2/16\pi G=8.0\times 10^{-30}$ g cm$^{-3}$.

Since the Planck length $l_{Pl}$ is a natural minimal length scale
in physics, we define the absolute minimal mass which possibly can
exist in nature by
\begin{equation}
M_{\min}=\frac{\Lambda c^{2}}{16\pi G} \frac{4\pi}{3}l_{Pl}^{3} =\frac{%
\Lambda c^{2}}{12G}l_{Pl}^{3} =\frac{\Lambda }{12}\sqrt{\frac{\hbar^{3}G}{%
c^{5}}} =\frac{\Lambda }{6\sqrt{2}}m_{Pl}l_{Pl}^{2} =\frac{\Lambda}{3}\frac{%
\hbar}{2}\frac{t_{Pl}}{2},  \label{minmass}
\end{equation}
where we denoted by $t_{Pl}$ the Planck time $t_{Pl}=l_{Pl}/c$.
The numerical value of $M_{\min}$ is given by
\begin{equation}
M_{\min} \approx 1.4 \times 10^{-127}\,\mathrm{g} \approx 7.9
\times 10^{-95}\,\mathrm{eV}.
\end{equation}

If an absolute minimum length does exist in nature, then, via the
first of Eqs.~(\ref{minm}), a positive cosmological constant
implies the existence of an absolute minimum mass in nature, given
by Eq.~(\ref{minmass}).

Hypothetical particles having this value of the mass may be called
cosminos. The cosminos could also be interpreted as ``quanta'' of the
dark energy (cosmological constant), and therefore $M_{\min}$
gives the mass of the quantum of the cosmological
constant. Compared with the upper bound of the electron neutrino
mass $m_{\nu_{e}} < 1.8\,\mathrm{eV}$~\cite{tr03}, we emphasize
the smallness of the minimal mass $M_{\min}$.

By generalizing Eq.~(\ref{minmass}) we propose that the mass is
quantized according to the general rule
\begin{equation}
m = n\, \frac{\Lambda }{3}\frac{\hbar }{2} \frac{t_{Pl}}{2},\qquad
n\in N,
\end{equation}
which is different from Wesson's proposal~\cite{We04}.

>From a purely quantum mechanical point of view, the value of the
minimum mass can be derived with the use of the uncertainty
principle for energy and time, which gives
\begin{equation}
m_{\min }c^{2}\approx \frac{\hbar }{\Delta t}.
\end{equation}

By assuming that $\Delta t$ is of the same order of magnitude as
the age of the Universe, $\Delta t\approx 1/H_0$, where
$H_0\approx 3.24\times 10^{-18}$ s$^{-1}$ is the Hubble constant
(the present value of the Hubble function), we obtain for the
minimum mass the expression
\begin{equation}
m_{\min }=\frac{\hbar H_0}{c^{2}}\approx 3. 8017\times
10^{-66}\mathrm{g}.
\end{equation}

The numerical value of the minimum mass obtained from quantum
mechanical
considerations agrees with the value of the mass $m_{P}=(\hbar/c) \sqrt{%
\Lambda /3}$ obtained by Wesson~\cite{We04} by using purely
dimensional considerations. Therefore it is natural to assume that
these two masses are the same, thus obtaining
\begin{equation}
\left(\frac{\hbar}{c}\right) \sqrt{\frac{\Lambda }{3}}=\frac{\hbar H_0}{c^{2}%
},
\end{equation}
which gives
\begin{equation}  \label{a}
H_0=c\sqrt{\frac{\Lambda }{3}}.
\end{equation}

We propose to call particles having the mass given by
$m_{\min}=m_P$ cosmons\footnote{Cosmons were originally introduced
by Peccei, Sola and Wetterich~\cite{Pe} to name scalar fields that
could dynamically adjust the cosmological constant to zero, see
also~\cite{So,We,So2}.}. The possibility of the existence of a
very light scalar particle, also named Cosmon, a dilaton which
should essentially decouple from the Standard Matter Lagrangian,
but still could mediate new macroscopic forces in the
submillimeter range, was proposed in~\cite{ShSo99}. The mass of
the Cosmon is given by $m_S^2=\Lambda_{QCD}^4/M^4$, where
$\Lambda_{QCD}\approx 100 \mathrm{MeV}$ is the intrinsic QCD scale
and $M\geq 10^{10} \mathrm{GeV}$ is some high energy
scale~\cite{ShSo99}. The mass of this particle is of the order of
the neutrino mass,
$m_S\approx(10^{-3}-10^{-2})\,\mathrm{eV}\approx 2\times
(10^{-30}-10^{-31})\,\mathrm{g}$. Therefore it represents a very
different particle as compared to the minimum mass particles
considered in the present paper.

By assuming however, that the minimum mass in nature is given by $%
m_{P}=m_{\min}$ it follows that the radius corresponding to
$m_{P}$ is given by
\begin{equation}
R_{P}=48^{1/6}\left( \frac{\hbar G}{c^{3}}\right)^{1/3}
\Lambda^{-1/6}\approx 1.9\,l_{Pl}^{2/3}\,\Lambda ^{-1/6},
\end{equation}
with the numerical value $R_P=4.7\times 10^{-13} \mathrm{cm} =4.7
\mathrm{fm} $. This would also imply that \emph{the minimum length
in nature could be very different from the Planck length
$l_{Pl}$}.

In fact the radius $R_P$ is of the same order of magnitude as the
classical radius of the electron $r_e=e^2/m_ec^2=2.81\times
10^{-13}$ cm. Therefore, by formally equating $R_P$ with $r_e$ and
neglecting terms of the order of unity gives a representation of
the \emph{cosmological constant in terms of the `classical'
fundamental constants} as
\begin{equation}\label{const}
\Lambda =\frac{l_{Pl}^{4}}{r_{e}^{6}}= \frac{\hbar ^{2}G^{2}m_{e}^{6}c^{6}}{%
e^{12}} \approx 1.4\times 10^{-56} \mathrm{cm}^{-2}.
\end{equation}

Conceptually, the identification of the radius $R_P$ to the electron radius $%
r_e$ may be based on a ``small number hypothesis'', representing
an extension of the large number hypothesis by Dirac~\cite{Di74},
and which proposes that \emph{the numerical equality between two
very small quantities with a very similar physical meaning cannot
be a simple coincidence}.

\section{Gravitational condensation of dark energy particles}

Recently a class of hypothetical compact objects called gravastars
({\it gra}vitational {\it va}cuum stars) have been proposed as
potential alternatives to explain the astrophysical phenomenology
traditionally associated to black holes \cite{MaMo04}. According
to this scenario, the quantum vacuum undergoes a phase transition
at or near the location the event horizon is expected to form.
Hence the gravastar consists of an interior de Sitter condensate,
governed by an equation of state $\rho c^2=-p$, matched to a shell
of finite thickness with an equation of state $\rho c^2=p$. The
latter is then matched to an exterior Schwarzschild solution. Dark
energy stars, for which the interior vacuum energy is much larger
than the cosmological energy, have also been investigated
\cite{Ch05,Lo06}. Hence the possibility that condensation
processes, like, for example, Bose-Einstein condensation, could
play an essential role in astrophysical and cosmological
situations cannot be excluded {\it a priori}.

Generally, Bose-Einstein condensation processes take place in a
Bose gas consisting of particles with mass $m$ and number density
$n$ when the thermal de Broglie wave length $\lambda
_{dB}=\sqrt{2\pi \hbar ^2/mkT}$, where $k$ is Boltzmann's constant
and $T$ is the temperature, exceeds the mean inter-particle
distance $n^{1/3}$, and the wave packets percolate in space. The
critical condensation temperature is $T\leq2\pi\hbar ^2n^{2/3}/mk$
\cite{BoHa07}. If we assume an adiabatic cosmological expansion of
the universe, the temperature dependence of the number density of
the particle is $T\propto n^{2/3}$. Hence Bose-Einstein
condensation occurs if the mass of the particle satisfies the
condition $m<1.87$ eV \cite{Bos}, a condition which is obviously
satisfied by both cosminos and cosmons. Hence these particles may
Bose-Einstein condense to form large scale astrophysical or
cosmological structures.

It is tempting to assume that the cosmons with mass $M_{\min }$ or
$m_{\min}=m_P$ could \emph{condense gravitationally} to form
stellar type stable compact objects. To study the cosmologival
implications of the condensation process we assume that the cosmon
fluid, with an initial density $\rho _{0}=\Lambda c^{2}/8\pi G$
and pressure $p_{0}$, satisfying the equation of state $\rho
_{0}c^{2}+p_{0}=0$, condenses into a non-relativistic,
dissipationless fluid, which can be characterized by a density
$\rho $, a pressure $p$, a velocity $\vec{v}$ and a gravitational
acceleration $\vec{g}$. The dynamics of the system is described by
the continuity equation, the hydrodynamical Euler equation and the
Poisson equation, which can be written as
\begin{equation}
\frac{\partial \rho }{\partial t}+\nabla \cdot \left( \rho \vec{v}\right) =0,\frac{%
\partial \vec{v}}{\partial t}+\left( \vec{v}\cdot \nabla \right) \vec{v}=-%
\frac{1}{\rho }\nabla p+\vec{g},  \label{hydr1}
\end{equation}
\begin{equation}
\nabla \times \vec{g}=0,\nabla \cdot \vec{g}=-4\pi G\rho .
\label{hydr2}
\end{equation}

We take as the initial (unperturbed) state of the system the state
characterized by the absence of the ''real'' gravitational forces, $\vec{g}=%
\vec{g}_{0}=0$, of the hydrodynamical flow,
$\vec{v}=\vec{v}_{0}=0$, and by
constant values of the density and pressure, $\rho =\rho _{0}$ and $p=p_{0}$%
, respectively, with $\rho _{0}c^{2}+p_{0}=0$. The condensation
process leads to the appearance of the gravitational interaction
in the system, as well as to small perturbations of the
hydrodynamical quantities, so that
\begin{equation}
\rho =\rho _{0}+\rho _{1},p=p_{0}+p_{1},\vec{v}=\vec{v}_{0}+\vec{v}_{1},\vec{%
g}=\vec{g}_{0}+\vec{g}_{1},
\end{equation}
so that $-1<<\rho _{1}/\rho _{0}<<1$ and $-1<<p_{1}/p_{0}<<1$,
respectively. In the first order approximation Eqs. (\ref{hydr1})
and (\ref{hydr2}) take the form
\begin{equation}
\frac{\partial \rho _{1}}{\partial t}+\rho _0\nabla \cdot
\vec{v}_{1}
=0,\frac{\partial \vec{v}_{1}}{\partial t}=-\frac{v_{s}^{2}}{\rho _{0}}%
\nabla \rho _{1}+\vec{g}_{1},  \label{hydr3}
\end{equation}
\begin{equation}
\nabla \times \vec{g}_{1}=0,\nabla \cdot \vec{g}_{1}=-4\pi G\rho
_{1},
\end{equation}
where we have introduced the adiabatic speed of sound $v_{s}$ in
the condensed cosmon fluid, defined as $v_{s}=\sqrt{p_{1}/\rho _{1}}=\sqrt{%
\partial p/\partial \rho }$. By taking the partial derivative with respect
to the time of the continuity equation in Eqs. (\ref{hydr3}), we
obtain the propagation equation of the density perturbation in the
cosmon fluid in the form
\begin{equation}
\frac{\partial ^{2}\rho _{1}}{\partial t^{2}}=v_{s}^{2}\nabla ^{2}\rho _{1}+%
\frac{\Lambda c^{2}}{2}\rho _{1}.
\end{equation}

By looking for a solution of the form $\rho _{1}\propto \exp
\left[ i\left( \vec{k}\cdot \vec{r}-\omega t\right) \right] $, we
obtain the following dispersion relation for $\omega $
\begin{equation}
\omega ^{2}=v_{s}^{2}\vec{k}^{2}-\frac{\Lambda c^{2}}{2}.
\end{equation}

>From the dispersion equation one can see that for $k<k_{J}$, where
\begin{equation}\label{kj}
k_{J}=\sqrt{\frac{\Lambda c^{2}}{2v_{s}^{2}}},
\end{equation}
is the Jeans wave number, the angular frequency $\omega $ becomes
an imaginary quantity, which corresponds to an instability of the
fluid-$\rho _{1}$can increase (or decrease) exponentially, leading
to
a gravitational condensation (or rarefaction). Therefore, for $k<k_{J}$, $%
\omega =\pm v_{s}\sqrt{k^{2}-k_{J}^{2}}=i{\rm Im}\omega $, where ${\rm Im}%
\omega =\pm v_{s}\sqrt{k_{J}^{2}-k^{2}}$, and consequently $\rho
_{1}\propto \exp \left[ \pm \left| {\rm Im}\omega \right| t\right]
$.

When the mass of the condensate exceeds the mass of a sphere with radius $%
2\pi /k_{J}$, a gravitational instability occurs in the cosmon
fluid, and
the cloud of particles would collapse . The critical mass is the Jeans mass $%
M_{J}=\left( 4\pi /3\right) \left( 2\pi /k_{J}\right) ^{3}\rho
_{0}$, and for the cosmon fluid it is given by
\begin{equation}
M_{J}=\frac{8\sqrt{2}}{3}\pi ^{3}\left( \frac{v_{s}}{c}\right) ^{3}\frac{%
c^{2}}{G}\Lambda ^{-1/2}\approx 1.6\times 10^{30}\times \left(
\Lambda \;{\rm cm}^{-2}\right) ^{-1/2}\left(
\frac{v_{s}}{c}\right) ^{3}\;{\rm g}.
\end{equation}

For $\Lambda =3\times 10^{-56}$cm$^{-2}$ we obtain $M_{J}=9.
24\times 10^{57}\left( v_{s}/c\right) ^{3}$ g. By taking into
account the representation of the cosmological constant in terms
of the ''classical constants'' given by Eq. (\ref{const}), we
obtain for the critical Jeans mass of the cosmon fluid the
expression
\begin{equation}
M_{J}=\frac{8\sqrt{2}}{3}\pi ^{3}\left( \frac{v_{s}}{c}\right) ^{3}\frac{%
e^{6}}{\hbar G^{2}m_{e}^{3}c}.
\end{equation}

The effective radius $R_{J}$ of the stable cosmon configuration is
given by
\begin{equation}
R_{J}=2^{3/2}\pi \frac{v_{s}}{c}\Lambda ^{-1/2}\approx 2^{3/2}\pi \frac{v_{s}%
}{c}\frac{e^{6}}{\hbar Gm_{e}^{3}c^{3}}.
\end{equation}

The theoretical value of the maximum mass $M_{Ch}$ of the stable
compact astrophysical type objects, like white dwarfs and neutron
stars, was found by Chandrasekhar and Landau and is given by the
Chandrasekhar limit,
\begin{equation}\label{chandra}
M_{Ch}\approx \left[ \left( \frac{\hbar c}{G}\right)
m_{B}^{-4/3}\right] ^{3/2},
\end{equation}
where $m_{B}$ is the mass of the particles giving the main
contribution to the mass (baryons in the case of the white dwarfs
and neutron stars)~\cite {ShTe83}. Thus, with the exception of
some composition-dependent numerical factors, the maximum mass of
a degenerate star depends only on fundamental physical constants.

The Jeans mass for cosmons can also be written in the form of a
Chandrasekhar limiting mass, if we assume that the cosmons have an
effective mass $m_{eff}$ given by
\begin{equation}
m_{eff}=\left( G\hbar ^{5}c^{5}\right)
^{1/4}\frac{m_{e}^{3/2}}{e^{3}},
\end{equation}
so that $M_{J}=\left[ 8\sqrt{2}\pi ^{3}\left( v_{s}/c\right) ^{3}/3\right] %
\left[ \left( \hbar G/c\right) m_{eff}^{-4/3}\right] ^{3/2}$. The
value of the effective mass of the cosmon is $m_{eff}\approx
8\times 10^{-20}$ g.

On the other hand, one can also assume that the cosminos or the
cosmons could form stellar type objects with the limiting mass
given by the Chandrasekhar limit, Eq. (\ref{chandra}. The mass of
such an hypothetical super-massive object formed from cosminos
with mass $M_{\min}$ is of the order of $M_{Ch}^{(1)}=8\times
10^{237}$ g, which exceeds by around $180$ orders of magnitude the
mass of our universe. Therefore, it follows that cosminos did not
condense gravitationally, and hence the particles associated with
dark energy fail to represent dark matter, which is in complete
agreement with the present standard model of cosmology. On the
other hand, in the case of cosmons for the Chandrasekhar limiting
mass $M_{Ch}^{(2)}$ we obtain $%
M_{Ch}^{(2)}=2.7782\times 10^{116}$ g, which also shows that
degenerate cosmon stars, having masses much larger than the mass
of the universe, are very unlikely to exist.

\section{Gravitational energy of stable cosmon configurations}

The total energy (including the gravitational field contribution)
inside an equipotential surface $S$ of radius $R$ can be defined,
according to~\cite {LyKa85}, to be
\begin{equation}
E=E_{M}+E_{F}=\frac{c^{4}}{8\pi G}\xi _{s}\int_{S}[K]dS,
\end{equation}
where $\xi ^{i}$ is a Killing vector field of time translation,
$\xi _{s}$ its value at $S$ and $[K]$ is the jump across the shell
of the trace of the
extrinsic curvature of $S$, considered as embedded in the 2-space $t=\mathrm{%
constant}$. $E_{M}=\int_{S}T_{i}^{k}\xi^{i}\sqrt{-g}dS_{k}$ and
$E_{F}$ are the energy of the matter and of the gravitational
field, respectively. This definition is manifestly coordinate
invariant.

For a static spherically symmetric system in a Schwarzschild-de
Sitter space-time the total energy is
\begin{equation}
E=\frac{c^{4}}{G}R\left[ 1-\left( 1-\frac{2GM}{c^{2}R}-\frac{\Lambda }{3}%
R^{2}\right) ^{1/2}\right] \left( 1-\frac{2GM}{c^{2}R}-\frac{\Lambda }{3}%
R^{2}\right) ^{1/2}.
\end{equation}

For the minimum mass particle the total energy can be expressed in
terms of the radius and cosmological constant only as
\begin{equation}  \label{en}
E=\frac{c^{4}}{G}R\left[ 1-\left( 1-\frac{\Lambda}{2}R^{2}\right)^{1/2}%
\right] \left( 1-\frac{\Lambda}{2}R^{2}\right) ^{1/2}.
\end{equation}
For a stable configuration, the energy should have a minimum,
$\partial E/\partial R=0$, a condition which determines $R$ as
\begin{equation}  \label{rad}
R_{BC}=\frac{1}{3}\sqrt{11+\sqrt{13}}\Lambda ^{-1/2} \approx
1.3\times \Lambda^{-1/2}.
\end{equation}
Therefore the mass of the stable configuration can be obtained as
\begin{equation}
M_{BC}=\frac{1.15}{6}\frac{c^{2}}{G}\Lambda
^{-1/2}\approx0.2\frac{c^{2}}{G}\frac{r_e^3}{l_{Pl}^2}\approx
0.2m_{Pl}\left(\frac{r_e}{l_{Pl}}\right)^3,
\end{equation}
which gives a mass of the order $M_{BC}\approx 8.2\times 10^{54}$
g, a value which is close to the mass $m_{PE}$, which follows from
dimensional considerations, and is of the same order of magnitude
as the total mass of the observable universe. Therefore we may
regard the observable universe as a dark energy dominated object
with minimum density.

For the second derivative of the energy, evaluated for $R=R_{BC}$,
we obtain the expression $\left.( \partial ^{2}E/\partial
R^{2})\right|_{R=R_{BC}} =-6.89\sqrt{\Lambda }$, which shows that
indeed the configuration is in a state of minimum total energy.

\section{Discussions and final remarks}

In the present paper we have investigated some of the possible
consequences of the existence of a minimum mass and density for
stable general relativistic objects, which is a direct result of
the existence of the cosmological constant. The existence of a
fundamental length, assumed to be the Planck scale, leads to an
absolute minimum mass in nature, which could be the mass of the
quanta of the dark energy (the cosminos), with radius of the order
of the Planck length. However, the application of the quantum
uncertainty principle for the energy shows that the mass of the
elementary particles associated to the dark energy (the cosmons)
is given by $m_{P}=\hbar H_0/c^2=\left( \hbar /c\right)
\sqrt{\Lambda /3}$. If this is indeed the case, then the radius of
such a particle is of the same order of magnitude as the classical
electron radius. This leads \emph{to the intriguing possibility of
the electron charge and radius, or, more generally, of the
electromagnetic processes, as playing an essential role in the
dark energy related phenomena}.

We also propose that ``dark energy particles'' may condensate,
either Bose-Einstein or gravitationally,  to form compact
super-massive objects, formed of cosminos or cosmons,
respectively. The mass of this condensation, which is
gravitationally stable, was derived using two independent methods.
Firstly, we have assumed that the dark energy fluid condenses into
a dissipationless, non-relativistic fluid. The corresponding Jeans
mass is proportional to $\Lambda ^{-1/2}$, and (except some
numerical factors) is the same as the mass $m_{PE}$ introduced
from dimensional considerations. Its numerical value is of the
same order of magnitude as the total mass of the universe.

The requirement that the total energy of the stable configuration
formed from the particle satisfying the relation $2GM= \Lambda /
6R^3$ is a minimum leads to a second, rigorous derivation of the
mass $m_{PE}$, which is of the same order of magnitude as the mass
of the universe. This also shows that the only energetically
stable dark energy dominated general relativistic objects must
have a mass of the same order of magnitude as our universe.
Therefore the general relativistic condition Eq.~(\ref{minm}) as
combined with the thermodynamic condition of energetic stability
may explain the actual value of the mass of the universe.
Moreover, the total mass of the universe can also be obtained in
terms of the elementary constants ${c,\hbar,e,m_e,G}$.

Therefore, these two independent results may imply that our
universe was born as the result of the dark energy condensation,
which took place at a very high temperature and density. Hence the
initial constituents of our universe may have been the cosmons. We
also obtain the physical interpretation of the masses $m_{PE}$ and
$M_{BC}$ as the critical Jeans mass of the Universe, that is, the
mass of the gravitationally stable dark particles clouds. This
result also gives a new physical interpretation of the
cosmological constant. From Eq. (\ref{kj}), by assuming that the
speed of sound in the gravitationally condensed dark energy fluis
is equal to the speed of light, $v_s=c$, it follows that $\Lambda
\approx k_J^2$, that is, \emph{physically the cosmological
constant represents the square of the Jeans wave number of a dark
energy fluid}. Alternatively, one can express the cosmological
constant as $\Lambda =8\pi^2/\lambda _J^2$, where $\lambda _J=2\pi
/k_J$ is the Jeans wavelength. Moreover, the mass of the universe
can be expressed in terms of three fundamental quantities, the
Planck mass $m_{Pl}$, the Planck length $l_{Pl}$, and the
classical electron radius $r_e$, respectively.

On the other hand, even that the estimation of the limiting
Chandrasekhar masses~\cite{ShTe83} for cosmons suggests the
possible existence of super-massive stable degenerate dark energy
objects, the existence of such stars with masses much larger than
the mass of the universe is impossible in the universe we are
living in.

Finally, it would be very interesting to recall the cosmological
constant problem again here. If it is interpreted as a measure of
the vacuum energy density and from a particles physics point of
view, the cosmological constant $\Lambda$ is 120 orders of
magnitude too small than expected~\cite {PeRa03}.

Let us therefore assume that the cosmological constant were indeed
120 orders of magnitude larger. This would have drastic
consequences for the minimal mass and we would find $M_{\min}
\approx 10^{19}\,\mathrm{eV}$, in which case the minimal mass
would exceed the masses of all elementary particles. From this
point of view, we would like to also argue that because of the
resulting problems, the interpretation of the cosmological
constant as the vacuum energy density may raise some conceptual
contradictions with the results derived in the present paper.

\subsection*{Acknowledgements}
We would like to thanks to the two anonymous referees, whose
comments helped us to significantly improve an earlier version of
the  manuscript. The work of TH was supported by the RGC grant
No.~7027/06P of the government of the Hong Kong SAR.


\begin{thebibliography}{}

\bibitem{Ri98} A. G. Riess et al., Astron. J. {\bf 116},
109 (1998)

\bibitem{Pe99} S. Perlmutter et al.,  Astrophys. J. {\bf 517}, 565
(1999)

\bibitem{Ber00} P. de Bernardis et al., Nature {\bf 404}, 995
(2000)

\bibitem{Ha00} S. Hanany et al., Astrophys. J. {\bf 545}, L5 (2000)

\bibitem{PeRa03} P. J. E. Peebles and B. Ratra,  Rev. Mod. Phys.
{\bf 75}, 559 (2003)

\bibitem{Pa03} T. Padmanabhan, Phys. Repts. {\bf 380}, 235 (2003)

\bibitem{MaDoHa00} M. K. Mak, P. N. Dobson, Jr. and T. Harko, Mod. Phys. Lett.
{\bf A 15}, 2153 (2000)

\bibitem{Bo04} C. G. B\"ohmer, Gen. Rel. Grav. {\bf 36}, 1039
(2004)

\bibitem{Bo05} C. G. B\"ohmer, Ukr. J. Phys. {\bf 50}, 1219 (2005)

\bibitem{BaBoNo05a} A. Balaguera-Antolinez, C. G. B\"ohmer and M. Nowakowski,
{\em Int. J. Mod. Phys.} D14, 1507 (2005)

\bibitem{Bu} H. A. Buchdahl, Phys. Rev. {\bf 116}, 1027 (1959)

\bibitem{BaBoNo05} A. Balaguera-Antolinez, C. G. B\"ohmer and M. Nowakowski,
  Class. Quant. Grav. {\bf 23}, 485 (2006)

\bibitem{We04}P. S. Wesson, Mod. Phys. Lett. {\bf A 19}, 1995 (2004)

\bibitem{Pe1} R. Aldrovandi, J. P. Beltran Almeida and J. G.
Pereira, arXiv:gr-qc/0702065 (2007)

\bibitem{Pe2} R. Aldrovandi, J. P. Beltran Almeida, C. S. O. Mayor and J. G.
Pereira, arXiv:0709.3947 (2007)

\bibitem{Pe3} R. Aldrovandi, J. P. Beltran Almeida, C. S. O. Mayor and J. G.
Pereira, arXiv:0710.0610 (2007)

\bibitem{Pe4} R. Aldrovandi, J. P. Beltran Almeida and J. G.
Pereira, Class. Quant. Grav. {\bf 24}, 1385 (2007)

\bibitem{BoHa05} C. G. B\"ohmer and T. Harko, Phys. Lett. {\bf B 630}, 73 (2005)

\bibitem{Bo1} C. G. B\"ohmer and T. Harko, Class. Quant. Grav. {\bf 23},
6479 (2006)

\bibitem{Bo2} C. G. B\"ohmer and T. Harko, Gen. Rel. Grav. {\bf 39},
757 (2007)

\bibitem{We89} S. Weinberg, Rev. Mod. Phys. {\bf 61}, 1 (1989)

\bibitem{BiDa82} N. D. Birrel and P. C. W. Davies, Quantum fields
in curved space, Cambridge University Press, Cambridge (1982)

\bibitem{BaFaHo07} A. B. Batista, J. C. Fabris and S. Houndjo,
arXiv:0710.0999 (2007)

\bibitem{We05} P. S. Wesson, Found. Phys. Lett. {\bf 19}, 285
(2006)

\bibitem{tr03} M. Trinczek et al., Phys. Rev. Lett. {\bf 90}, 012501 (2003)

\bibitem{Pe} R. D. Peccei, J. Sola and C. Wetterich, Phys. Lett. {\bf B 195}, 183 (1987)

\bibitem{We} C. Wetterich, Nucl. Phys. {\bf B 302}, 668 (1988)

\bibitem{So} J. Sola, Phys. Lett. {\bf B 228}, 317 (1989)

\bibitem{So2} J. Sola,  Int. J. Mod. Phys. {\bf A 5}, 4225 (1990)

\bibitem{ShSo99} I. L. Shapiro and J. Sola, Phys. Lett. {\bf B 475}, 236 (2000)

\bibitem{Di74} P. A. M. Dirac,  Proc. R. Soc. {\bf A 333}, 439 (1974)

\bibitem{MaMo04} P. O. Mazur and E. Mottola, Proc. Nat. Acad. Sci. {\bf 101},
9545 (2004)

\bibitem{Ch05} G. Chapline, arXiv:astro-ph/0503200 (2005)

\bibitem{Lo06} F. S. N. Lobo, Class. Quant. Grav. {\bf 23}, 1525
(2006)

\bibitem{BoHa07} C. G. B\"ohmer and T. Harko, JCAP {\bf 06} 025
(2007)

\bibitem{Bos} T. Fukuyama, M. Morikawa and T. Tatekawa,
arXiv:0705.3091 (2007)

\bibitem{ShTe83} S. L. Shapiro and S. A. Teukolsky, Black Holes, White Dwarfs, and Neutron Stars, John Wiley \& Sons, New York (1983)

\bibitem{LyKa85} J. Katz, D. Lynden-Bell and W. Israel, Class. Quantum Grav. {\bf 5}, 971 (1988)

\end{thebibliography}
\end{document}